\documentclass[journal,12pt,draftclsnofoot,onecolumn,doublespace]{IEEEtran}
\usepackage{amsmath,amssymb,epsfig,psfrag,cite,subfigure}
\include{macros}
\usepackage{graphicx}
\usepackage{algorithm}
\usepackage{epsfig,psfrag}
\usepackage{subfigure}
\usepackage{color}
\usepackage{url}

\usepackage{graphicx}

\begin{document}

\title{TW-TOA Based Positioning in the Presence of Clock Imperfections }
\author{{Mohammad~Reza~Gholami, \emph{Member, IEEE},\\Sinan~Gezici, \emph{Senior Member, IEEE}, and~Erik~G.~Str\"om, \emph{Senior Member, IEEE}}
\thanks{M. R. Gholami is with the ACCESS Linnaeus Center, Electrical Engineering, KTH--Royal Institute of Technology, SE-100 44 Stockholm, Sweden
(e-mail: mohrg@kth.se)}
\thanks{S. Gezici is with the Department of Electrical and Electronics Engineering, Bilkent University, Ankara 06800, Turkey (e-mail: gezici@ee.bilkent.edu.tr).}
\thanks{E. G. Str\"om is with the Division of Communication Systems, Information Theory, and Antennas, Department of Signals and Systems, Chalmers University of Technology, SE-412 96 Gothenburg, Sweden (e-mail: erik.strom@chalmers.se).}
\thanks{This work was supported in part by
  the Swedish Research Council (contract no.~2007-6363) and in part by the European Commission
 in the framework of the FP7 Network of Excellence in Wireless
  COMmunications COMmunications \#
(contract no. 318306).  Part of this work was presented in IEEE ICASSP 2013~\cite{gholami2013range}.}}

\maketitle

\IEEEpeerreviewmaketitle

\begin{abstract}
This paper studies the positioning problem based on two-way time-of-arrival (TW-TOA) measurements
in asynchronous wireless sensor networks. Since the optimal estimator for this problem involves difficult nonconvex optimization, we propose two suboptimal estimators based on squared-range least squares and
least absolute mean of residual errors. The former approach is formulated as a general trust region subproblem which
can be solved exactly under mild conditions.
The latter approach is formulated as a \emph{difference of convex functions programming} (DCP),
which can be solved using a concave-convex procedure. Simulation results illustrate the high performance of the proposed techniques, especially for the DCP approach.

\end{abstract}

\section{Introduction}
Location aware services are becoming vital requirements for many wireless systems.
Due to some drawbacks of using GPS receivers at  wireless nodes for some scenarios, self-position recovery
has been proposed as an alternative approach and extensively investigated in the literature~\cite{dardari2011satellite,Localization_algorithm_2009,Patrwari_2005_Locating_Nodes,Sinan_2005,Sayed_2005,Lic_Mohammad}.
 Positioning based on range estimates between nodes is a popular technique in the literature. For synchronous networks, the time-of-arrival (TOA) technique provides a good estimate of the distance between two nodes for reasonable signal-to-noise ratios.
A huge number of algorithms have been proposed in the literature
to address the positioning problem based on range measurements, e.g., the maximum likelihood estimator~\cite{Localization_algorithm_2009},
linear least-squares~\cite{Wang_2003_correctionTechnique,Ho_2009_Succesive_Asymptotically,Gholami_ICC2011}, squared-range least squares~\cite{Amir_LS_2008},
projection onto convex sets~\cite{Hero_POCS_2005,Gholami_Eurasip_2011,Gholami_PIMRC_2010}, and convex relaxation techniques~\cite{SOCP_Tseng_2007,SDP_Biswas_2006,Sum_of_squares}.

In asynchronous networks,
the range estimate based on the TOA is highly sensitive to the clock imperfections. Therefore, the positioning accuracy can be considerably degraded in the presence of clock imperfection. In particular for an affine model describing  the clock behavior, the accuracy of  the positioning techniques based on the TOA measurements is affected by non-ideal clock offset and  clock skew.
The clock of a target node can be synchronized with a reference time (clock) using a synchronization technique using, e.g., the MAC layer time stamp exchanging, e.g., see,\cite{Erchin_2009_book,Chaudhari_clock_2008,wanlu_2012,Chaudhari_clock_2010_IT,chaudhari_2011_clocksynch} and references therein.  Motivated by pairwise synchronization techniques, the authors in~\cite{Joint_syn_local_2010} formulate
a joint synchronization and positioning problem in the MAC layer. If
the major delay is the fixed delay due to propagation through the
radio channel, the joint position and timing estimation technique
works well. The authors in \cite{cheng2004tps} study the positioning problem in the presence of clock imperfection only due to clock offset. Considering the effect of imperfect clock on distance estimates in the physical layer, the authors in \cite{TDOA_gholami_2013} investigate the positioning problem using time-difference-of-arrival (TDOA) in the presence of clock imperfections. The TDOA technique effectively removes the clock offset, but still suffers from the clock skew.

Another popular approach for estimating the distance between sensor nodes is to use a so-called  two-way time-of-arrival (TW-TOA)  or time-of-flight based technique, which is an elegant approach in removing the effect of the clock offset on range measurements~\cite{Gholami_ICC2011}.
Range estimates obtained via TW-TOA are affected by the clock skew and a processing delay called the turn-around time~\cite{Gholami_SPAWC_2011_1}.
A number of researchers have tackled the positioning problem or distance estimation based on TW-TOA in fully or partially asynchronous
networks~\cite{Gholami_Coop_2012,RobustClock_Localization_2011}.
To improve the range estimation via TW-TOA, an effective technique based on
a new clock counter scheme is proposed in~\cite{Sahinoglu_2011}.
The authors in \cite{RobustClock_Localization_2011} study the positioning problem in the presence of clock imperfections for a TW-TOA based technique and propose a linear least squares based approach to solve the problem.
The proposed approaches work well in some scenarios, e.g., when there is a sufficient number of reference nodes at known positions.
In general, the previously proposed approaches need modifications
to be effectively applied to the positioning problem in which  the clock skew and turn around times are also unknown.
In addition, for practical applications the proposed algorithms may not be robust against outliers and non-line-of-sight errors.

In this study, we consider the positioning  of a single target node based on TW-TOA measurements in the presence of clock imperfections.
In this approach, a target node transmits
a signal to a reference node located at a known position and the reference node responds to the received signal after an unknown turn-around time delay.
As it is common in the literature, we assume that the reference node measures the turn-around time
by a loop back test and transmits the estimate to the
target node~\cite{Gezici_ICC10,Sahinoglu_2011}. The target node then computes the round-trip delay based on an estimate of the turn-around time.
Assuming an affine model for the clock of the oscillator, it is observed that the range estimation using
the TW-TOA measurement is affected by an unknown clock skew of the target node. Modeling the measurement errors as Gaussian random variables, we obtain the optimal estimator to find the clock skew, and the location of the target node, and the turn-around times for the reference nodes. The optimal estimator poses a high dimensional optimization problem and needs more than one distance estimate for every link to provide good estimates of the unknown parameters. We, then, omit the effect of the turn-around times using a linear transformation and consequently obtain a near-optimal estimator to find the location and clock skew of the target node.
Both the optimal and near-optimal estimators for the positioning problem considered in this study are nonconvex and difficult to solve.
Using some approximations, we obtain two suboptimal estimators. In the first approach, we apply a
similar technique as considered in~\cite{Amir_LS_2008} for synchronized networks to the
asynchronous scenario. Namely, we consider the squared-range least-squares approach and formulate the problem as a general trust
region subproblem, which can be solved exactly under mild conditions.
In the second approach, we minimize the residual errors based on the $\ell_1$ norm and arrive at a nonconvex problem in the
form of the \emph{difference of convex functions programming} (DCP).
We then employ a concave-convex procedure to solve the problem. Note that the latter approach is robust against outliers.
Simulation results indicate the high performance of the proposed techniques, especially the DCP.

In summary the \emph{main contributions} of this study for TW-TOA based positioning in the presence of clock imperfections are:
\begin{itemize}
\item the MLE and an approximate MLE to find the location and clock skew of the target node;
\item a suboptimal estimator based on general trust region subproblem for squared-range measurements;
\item a suboptimal estimator formulated as the DCP that can be solved using a concave-convex procedure.
\end{itemize}

The remainder of the paper is organized as follows.
Section~\ref{sec:signal_model} explains the signal model considered in this study.
In Section~\ref{sec:MLE} and~\ref{sec:subOptEsts} the localization algorithms are studied.
Complexity analyses of different approaches are discussed in Section~\ref{Sec:complexiy}.
Simulation results are presented
in Section~\ref{sec:simulation}.
Finally, Section~\ref{sec:conclude} makes come concluding remarks.

\textbf{Notation}: The following notations are used in this paper. Lowercase Latin/Greek letters, e.g., $a,b,\beta$, denote scalar values and
bold lowercase Latin/Greek letters represent vectors. Matrices are shown by bold uppercase Latin/Greek letters.  $\textbf{I}_{M}$ is the $M$ by $M$ identity matrix.
The operator $\mathbb{E}\{\cdot\}$ is used to denote
the expectation of a random variable (or vector). The $\ell_p$ norm of a vector is denoted
by $\|\cdot\|_p$.  The $\text{diag}(X_1,\ldots,X_N)$ is a diagonal matrix with diagonal elements $X_1,\ldots,X_N$.
For two matrices $\boldsymbol{A}$ and $\boldsymbol{B}$, $\boldsymbol{A} \succeq \boldsymbol{B}$ means $\boldsymbol{A}-\boldsymbol{B}$ is positive semidefinite.
$\bigtriangledown g(\boldsymbol{a})$ denotes the gradient of $g(\boldsymbol{x})$
at $\boldsymbol{x}=\boldsymbol{a}$. The set of all $N$-vector with positive components are denoted by $\mathbb{R}_+^N$. We use $\otimes$ to denote the Kronecker product.

\section{System Model}
\label{sec:signal_model}
Consider a two dimensional network\footnote{The generalization to a three dimensional network is straightforward, but is not explored in this study.}
with $N$ reference (anchor) nodes located at known
positions ${\boldsymbol{a}_i=[a_{i,1}~a_{i,2}]^T\in\mathbb{R}^2}$, $i=1,...,N$.  Suppose that one target node
is placed at unknown position $\boldsymbol{x}=[x_1~x_2]^T\in \mathbb{R}^2$.
We assume that the target node estimates the distance to a reference node by performing a TW-TOA measurement.
We further assume that the clock value of an imperfect clock follows an affine
relation with the true (global) time $t$~\cite{Davide_ranging_2009,Erchin_2009_book,Chaudhari_clock_2008,Chaudhari_clock_2010_IT,chaudhari_2011_clocksynch}. That is, the clock
value of reference node $i$ is
\begin{equation}
  \label{eq:1}
  g_i(t) \triangleq w_i t + \theta_i
\end{equation}
where $w_i$ is the skew and $\theta_i$ is the offset associated with
the $i$th node clock. Note that a perfectly synchronized clock has
$w_i = 1$ and $\theta_i = 0$. In practice, $w_i$ is a number close to
1. For convenience, we denote the target clock as $g(t)$, where $g(t)
= wt + \theta$.

A TW-TOA measurement between the target node and the $i$th reference
node is carried out as follows: (a) the target sends a message to the
reference node at (global) time $t^k_{i,1}$, (b) the message arrives at
the reference node at time $t^k_{i,2}$, (c) the reference node sends a
return message at time $t^k_{i,3}$, and (d) the return message arrives
at the target node at time $t^k_{i, 4}$. Clearly, $t^k_{i,2} - t^k_{i,1} =
t^k_{i, 4} - t^k_{i, 3} = d_i/c$, where $c$ is the propagation speed and
$d_i \triangleq d(\boldsymbol{x}, \boldsymbol{a}_i) \triangleq
\|\boldsymbol{x}-\boldsymbol{a}_i\|_2$ is the distance between the target and
$i$th reference node. Moreover, $t^k_{i,3} = t^k_{i, 2} + T_i$, where
$T_i$ is the turn-around time in the $i$th reference node, which is assumed to be fixed during the positioning process. The
TWO-TOA measurement is computed in the target local clock
as
 \begin{equation}
  \label{eq:3}
  z^k_i = \frac{1}{2}\big[g(t^k_{4, i}) - g(t^k_{1, i}) + n^k_i \big] = w \frac{d_i}{c} + w\frac{T_i}{2} + \frac{n^k_i}{2},~k=1,2,\ldots, K
\end{equation}
where $n^k_i$ is the TW-TOA measurement error, which we model it as a
zero-mean Gaussian with standard deviation $\sigma_i$, i.e., $n^k_i \sim{\cal N}(0, \sigma_i^2)$, and $K$ as the number of the TW-TOA measurements during the positioning process.

The unknown parameter $T_i$ either might be extremely small and can be
neglected [20] or it needs to be estimated.  One way to deal with the
unknown parameter $T_i$ is to jointly estimate it along with the
location of the target node \cite{Gholami_SPAWC_2011}. It can also be estimated by
reference node $i$ using a loop back test and is sent back to the
target node \cite{Sahinoglu_2011}. In this study, we consider the latter approach. The
estimate of $T_i$ is
\begin{equation}
  \label{eq:2}
  \tilde{T}_{i}^k= g_i(t^k_{3, i}) - g_i(t^k_{2, i}) + \epsilon^k_i = w_i T_i +  \epsilon^k_i,~k=1,2,\ldots, K
\end{equation}
where we model the estimation error as $\epsilon\sim\mathcal{N}(0, \gamma_i^2)$.

In the sequel, we assume that the reference nodes are synchronized
with a reference clock, e.g., via a GPS signal.
Therefore $w_i \approx 1$  and we can write $\tilde{T}_i^k \approx \hat T_i^k $, where
\begin{equation}
  \label{eq:turn}
  \hat T_i^k \triangleq T_i + \epsilon^k_i.
\end{equation}
We now combine  \eqref{eq:turn} with~(\ref{eq:3}) and obtain
\begin{equation}
  \label{eq:approx_model}
  z^k_i - w\frac{\hat T_i^k}{2} = w \frac{d_i}{c} + \frac{n^k_i}{2} -w\frac{\epsilon^k_i}{2}.
\end{equation}
As mentioned, the approximation is good in the (reasonable) case when the reference nodes are equipped with accurate clocks.

In the following sections, we use the input data $\big\{\{z^k_i,
\hat{T}_i^k\}_{i=1}^N\big\}_{k=1}^K$ to obtain the optimal estimator based on models  \eqref{eq:3}-\eqref{eq:turn}  or suboptimal estimators according to \eqref{eq:approx_model}. The parameters $w$
and $T_i$, $i = 1, 2, \ldots, N$, are considered as unknown nuisance
parameters, while $\sigma_i$, $\gamma_i$, and $\boldsymbol{a}_i$ are
assumed to be known for $i = 1, 2, \ldots, N$.

\section{Maximum Likelihood Estimator}
\label{sec:MLE}
We define the measurement vector
\begin{align}
  \label{eq:meas}
  \boldsymbol{m} \triangleq
  \left[\begin{array}{c}\boldsymbol{m}^1\\
                                    \boldsymbol{m}^2\\
                                    \vdots\\
                                    \boldsymbol{m}^K \end{array}\right],
\end{align}
where
\begin{equation}
  \boldsymbol{m}^k \triangleq
  \begin{bmatrix}
    z^k_1 & z^k_2 & \cdots & z^k_N & \hat T_1^k & \hat T_2^k
    & \cdots & \hat T_N^k
  \end{bmatrix}^T.
\end{equation}
To obtain the MLE for joint estimation of the position and clock skew of the target node, the
following optimization problem needs to be solved~\cite{Kay_93}:
\begin{equation}
  \label{eq:8}
\Big[\hat w ~ \hat{\boldsymbol{t}}^T_a~ \hat{\boldsymbol{x}}^T \Big]
  =\mathop{\mathrm{arg~ max}}\limits_{w\in \mathbb{R}_+;\;\boldsymbol{t}_a\in\boldsymbol{R}_+^N;\, \boldsymbol{x}\in \mathbb{R}^2}  p(\boldsymbol{m}; w,\boldsymbol{t}_a, \boldsymbol{x})
\end{equation}
where $p(\boldsymbol{m};w,\boldsymbol{t}_a,\boldsymbol{x})$ is the probability
density function (pdf) of vector $\boldsymbol{m}$ indexed by
the vector $[w~\boldsymbol{t}^T_a~\boldsymbol{x}^T]^T$ and $\boldsymbol{t}_a=[T_1~T_2 \ldots T_N]^T$. Since the TOA measurement errors are assumed to be independent and identically distributed
random variables, the pdf of $\boldsymbol{m}$ can be calculated from \eqref{eq:3} and \eqref{eq:turn} as
\begin{align}
  \label{eq:pdf}
 & p(\boldsymbol{m}; w,\boldsymbol{t}_a, \boldsymbol{x}) =
  \prod_{k=1}^K\prod_{i=1}^N \sqrt{\frac{2}{\pi \sigma_i^2 }}
  \exp\left(- \frac{2(z^k_i - w T_i/2 -w d(\boldsymbol{x}, \boldsymbol{a}_i)/c)^2}{\sigma_i^2}\right) \sqrt{\frac{1}{2\pi \gamma_i^2}}
  \exp\left(- \frac{(\hat{T}_i^k -  T_i)^2}{2\gamma_i^2}\right).
\end{align}
Then, the MLE is obtained as
\begin{align}
\label{eq:MLE1}
\left[\hat{\boldsymbol{x}}^T~\hat{w}~~\hat{\boldsymbol{t}}_a \right]^T&=\mathop{\mathrm{arg~max}}\limits_{w\in \mathbb{R}_+;\, \boldsymbol{x}\in \mathbb{R}^2;\,T_i\in\mathbb{R}_+}  p(\boldsymbol{m}; w,\boldsymbol{t}_a, \boldsymbol{x})\nonumber\\
&=\mathop{\mathrm{arg~min}}
\limits_{\boldsymbol{x}\in\mathbb{R}^2;\,w\in\mathbb{R}_+;\,T_i\in\mathbb{R}_+}
\sum_{k=1}^K \sum_{i=1}^N \frac{2}{\sigma_i^2}\Big(z^k_i-w\frac{{T}_i}{2}-w\frac{d_{i}}{c}\Big)^2+\frac{(\hat{T}_i^k-T_i)^2}{\gamma_i^2}.
\end{align}

For the MLE formulated in \eqref{eq:MLE1} there are $N+3$ unknowns. Therefore, for low numbers of messages $K$, the MLE problem can be ill-posed. To alleviate the difficulty for solving the optimal MLE in \eqref{eq:MLE1}, we here investigate another approximate MLE based on the model obtained in \eqref{eq:approx_model}.

Let us define a new set of measurements as
\begin{align}
\label{eq:transformed_set}
   y^k_i\triangleq z^k_i - w\frac{\hat T_i^k}{2} = w\frac{d_i}{c} + \frac{n^k_i}{2} -w\frac{\epsilon^k_i}{2},\quad &i=1,2,\ldots,N,\nonumber\\&k=1,2,\ldots,K.
\end{align}
Now, we collect $y^k_i$ in a vector $\boldsymbol{y}=[y^1_1~\ldots~y^1_N~\ldots~y^K_1~\ldots~y^K_N ]^T$.  Next, we compute the pdf of
$\boldsymbol{y}$ as
\begin{align}
\label{eq:pdf}
p(\boldsymbol{y};w,\boldsymbol{x})
&=\prod_{k=1}^K\prod_{i=1}^N\sqrt{\frac{2}{\pi(\sigma_i^2+w^2\gamma_i^2)}}\exp{\left(-\frac{2(z^k_i- w{d_i}/{c}-w{\hat{T}_i^k}/{2})^2}{(\sigma_i^2+w^2\gamma_i^2)}\right)}.
\end{align}
We then find an approximate MLE (AMLE)\footnote{We call the MLE in \eqref{eq:MLE2} as AMLE because it is based on the transformed set of measurements in \eqref{eq:transformed_set} instead of the original measurements in \eqref{eq:meas}.} as
\begin{align}
\label{eq:MLE2}
\left[\hat{\boldsymbol{x}}^T~\hat{w}\right]^T&=\mathop{\mathrm{arg~max}}\limits_{w\in \mathbb{R}_+;\, \boldsymbol{x}\in \mathbb{R}^2}  p(\boldsymbol{z}; w, \boldsymbol{x})\nonumber\\
&=\mathop{\mathrm{argmin}}\limits_{\boldsymbol{x}\in\mathbb{R}^2;\, w\in\mathbb{R}_+}
\sum_{k=1}^K\sum_{i=1}^N \frac{2}{(\sigma_i^2+w^2\gamma_i^2)}\Big(z^k_i-w\frac{\hat{T}_i^k}{2}-w\frac{d_{i}}{c}\Big)^2
+{\ln(\sigma_i^2+w^2\gamma_i^2)}.
\end{align}
It is observed that the search domain in the AMLE in \eqref{eq:MLE2} is limited to the location $\boldsymbol{x}$ and the clock skew $w$, thus a lower dimensional search compared to that of the MLE in \eqref{eq:MLE1}.

It is also noted that both the MLE and AMLE formulations in \eqref{eq:MLE1} and \eqref{eq:MLE2} pose difficult global optimization problems. To avoid the drawbacks in solving these problems, we propose two suboptimal estimators in the next section.

\section{Proposed techniques}
\label{sec:subOptEsts}
In this section, we propose two techniques based on squared-range least squares and
$\ell_1$ norm minimization of residuals. First, we divide both sides of~\eqref{eq:approx_model}
by $w$ (we safely assume that $w\neq 0$) and express the model as
\begin{align}
\label{eq:sub_model}
z^k_i\alpha-\frac{\hat{T}_i^k}{2}= \frac{d_i}{c}+\frac{n^k_i}{2}\alpha-\frac{\epsilon^k_i}{2},~i=1,2,\ldots,N, ~k=1,2,\ldots,K
\end{align}
where $\alpha=1/w$.

In the following, the model in~\eqref{eq:sub_model} is employed in order to derive the proposed suboptimal estimators.

\subsection{Squared-Range measurement Least Squares (SQ-LS)}
\label{eq:sr-ls}
In this section, we assume that the measurement noise ${n^k_i}/{2}\alpha-{\epsilon^k_i}/{2}$ is small compared to $d_i/c$. Then, taking the square of both sides of \eqref{eq:sub_model}  and dropping the small terms yield
\begin{align}
\label{eq:sq-range}
(z^k_i\alpha)^2+\frac{(\hat{T}^k_i)^2}{4}-z^k_i\hat{T}_i^k\alpha&\simeq
\frac{1}{c^2}(\boldsymbol{x}^T\boldsymbol{x}-2\boldsymbol{a}_i^T\boldsymbol{x}+\|\boldsymbol{a}_i\|_2^2)+\nu^k_i,
\end{align}
where $\nu^k_i=d_i(\alpha n^k_i-\epsilon^k_i)/c$.
Now, we apply a weighted least squares criterion to the model in \eqref{eq:sq-range} and obtain the following minimization problem:
\begin{align}
\label{eq:sq-range2}
&\mathop{\mathrm{minimize}}\limits_{\boldsymbol{x}\in\mathbb{R}^2;~\alpha\in\mathbb{R}_+}~
\sum_{k=1}^K\sum_{i=1}^N\frac{1}{d_i^2(\alpha^2\sigma_i^2+\gamma_i^2)}\Bigg(\frac{1}{c^2}\boldsymbol{x}^T\boldsymbol{x}
-\frac{2}{c^2}\boldsymbol{a}_i^T\boldsymbol{x}-(z^k_i)^2\alpha^2+z^k_i\hat{T}_i^k\alpha+\frac{1}{c^2}\|\boldsymbol{a}_i\|_2^2-\frac{(\hat{T}_i^k)^2}{4}\Bigg)^2.
\end{align}
The problem in~\eqref{eq:sq-range2} can be expressed as a quadratic programming problem:
\begin{align}
\label{eq:qp}
&\mathop{\text{minimize}}\limits_{\boldsymbol{y}}~~ \|\boldsymbol{W}^{1/2}(\boldsymbol{A}\boldsymbol{y}-\boldsymbol{b})\|_2^2\nonumber\\
&\text{subject to}~\boldsymbol{y}^T\boldsymbol{D}\boldsymbol{y}+2\boldsymbol{f}^T\boldsymbol{y}=0
\end{align}
where  matrices $\boldsymbol{W}$, $\boldsymbol{A}$, and $\boldsymbol{D}$ and vectors $\boldsymbol{b}$, $\boldsymbol{f}$, and $\boldsymbol{y}$ are defined as
\begin{align}
\label{eq:linearMod_parameters}
&\boldsymbol{W}= \boldsymbol{I}_K\otimes\mathrm{diag}\left(\frac{1}{d^2_1(\alpha^2\sigma_1^2+\gamma_1^2)},\ldots,\frac{1}{d^2_N(\alpha^2\sigma_N^2+\gamma_N^2)}\right),\nonumber\\
&\boldsymbol{A}\triangleq\left[\begin{array}{cccc}
\frac{1}{c^2}&-\frac{2}{c^2}\boldsymbol{a}^T_1&-(z^1_1)^2&z^1_1\hat{T}_1^1\\
\vdots&\vdots&\vdots&\vdots\\
\frac{1}{c^2}&-\frac{2}{c^2}\boldsymbol{a}^T_N&-(z^1_N)^2&z^1_N\hat{T}_N^1\\
\vdots&\vdots&\vdots&\vdots\\
\frac{1}{c^2}&-\frac{2}{c^2}\boldsymbol{a}^T_1&-(z^K_1)^2&z^K_1\hat{T}_1^K\\
\vdots&\vdots&\vdots&\vdots\\
\frac{1}{c^2}&-\frac{2}{c^2}\boldsymbol{a}^T_N&-(z^K_N)^2&z^K_N\hat{T}_N^K
\end{array}\right],\nonumber\\
&\textbf{b}\triangleq\left[\begin{array}{ccc}
-\frac{1}{c^2}\|\boldsymbol{a}_1\|_2^2+\frac{(\hat{T}_1^1)^2}{4}\\
\vdots\\
-\frac{1}{c^2}\|\boldsymbol{a}_N\|_2^2+\frac{(\hat{T}_N^1)^2}{4}\\
\vdots\\
-\frac{1}{c^2}\|\boldsymbol{a}_1\|_2^2+\frac{(\hat{T}_1^K)^2}{4}\\
\vdots\\
-\frac{1}{c^2}\|\boldsymbol{a}_N\|_2^2+\frac{(\hat{T}_N^K)^2}{4}
\end{array}\right],\nonumber\\
&\boldsymbol{D}\triangleq \boldsymbol{I}_K\otimes\text{diag}(0,1,1,0,1),\nonumber\\
&\boldsymbol{f}\triangleq\boldsymbol{1}_K\otimes \left[-\frac{1}{2}~~{0}~~0~-\frac{1}{2}~~0\right]^T,\nonumber\\
&\boldsymbol{y} \triangleq \Big[\|\boldsymbol{x}\|_2^2~\boldsymbol{x}^T~\alpha^2~\alpha\Big]^T.
\end{align}
with $\boldsymbol{1}_k$ denoting a column vector of $k$ ones.

The problem in~\eqref{eq:qp} minimizes a quadratic function over a quadratic constraint. This type of problems are called a
generalized trust
region subproblem~\cite{Trust_region_93} and can be solved exactly.
It
has also been known that the general trust region subproblem has zero duality gap and the optimal solution can be extracted from the dual solution~\cite{GTR_New_82,Beck_GTR,Trust_region_93}. A necessary and sufficient condition for $\boldsymbol{y}^*$ to be optimal in \eqref{eq:qp} is that there exists a $\mu\in\mathbb{R}$ such that \cite{GTR_New_82}
\begin{align}
\label{eq:cond}
&(\boldsymbol{A}^T\boldsymbol{W}\boldsymbol{A}+\mu\boldsymbol{D})\boldsymbol{y}^*
=(\boldsymbol{A}^T\boldsymbol{W}\boldsymbol{b}-\mu\boldsymbol{f}),\nonumber\\
&(\boldsymbol{y}^*)^T\boldsymbol{D}\boldsymbol{y}^*+
2\boldsymbol{f}^T\boldsymbol{y}^*=0,\nonumber\\
&(\boldsymbol{A}^T\boldsymbol{W}\boldsymbol{A}+\mu\boldsymbol{D})\succeq 0.
\end{align}
Under the conditions considered in \eqref{eq:cond}, the solution to the problem  of \eqref{eq:qp} is given by
\begin{align}
\label{eq:op_sol}
\boldsymbol{y}(\mu)=(\boldsymbol{A}^T\boldsymbol{W}\boldsymbol{A}+
\mu\boldsymbol{D})^{-1}(\boldsymbol{A}^T\boldsymbol{W}\boldsymbol{b}-\mu\boldsymbol{f}).
\end{align}
In such a situation to find $\mu$, we simply replace \eqref{eq:op_sol} into constraint $\boldsymbol{y}^T\boldsymbol{D}\boldsymbol{y}+2\boldsymbol{f}^T\boldsymbol{y}=0$, i.e.,
\begin{align}
\label{eq:gamm}
\phi(\mu)= {\boldsymbol{y}^T}(\mu)\boldsymbol{D}_1{\boldsymbol{y}^T}(\mu)+2{\boldsymbol{f}^T}{\boldsymbol{y}}(\mu)=0,\quad \mu\in\mathcal{I}
\end{align}
%
where the interval $\mathcal{I}$ consists of all $\mu$ such that $\boldsymbol{A}^T\boldsymbol{W}\boldsymbol{A}+\mu\boldsymbol{D}\succeq~0$.
The interval $\mathcal{I}$ is given by~\cite{Amir_LS_2008}
\begin{align}
\label{eq:gamma_op}
\mathcal{I}=(-{1}/{\mu_1},\infty),
\end{align}
with $\mu_1$ representing the largest eigenvalue of $(\textbf{A}^T\boldsymbol{W}\textbf{A})^{-1/2}\boldsymbol{D}(\textbf{A}^T\boldsymbol{W}\textbf{A})^{-1/2}$~\cite{Trust_region_93}.
In summary, the solution to \eqref{eq:qp} is obtained as follows:
\begin{itemize}
\item Use a bisection search to find a root of $\phi(\mu)=0$, say $\mu^*$. Note that $\phi(\mu)$ is a strictly decreasing function with respect to
$\mu$ \cite{Trust_region_93}.
\item Replace $\mu^*$ into \eqref{eq:op_sol} to obtain $\boldsymbol{y}^*=\boldsymbol{y}(\mu^*)$.
\item Estimate the unknown parameters as $\hat{\boldsymbol{x}}=[\boldsymbol{y}^*]_{2:3}$ and~$\hat{w}=1/[\boldsymbol{y}^*]_4$, with $[\boldsymbol{v}]_{i:j}$ denoting the $i$th to the $j$th elements of vector $\boldsymbol{v}$.

\end{itemize}

Note that since the weighting matrix $\boldsymbol{W}$ depends on the unknown distance $d_i$ and $\alpha$, we first replace $\boldsymbol{W}$ with the identity matrix and find an estimate of the location and $\alpha$ as described above. Then, we reconstruct the distance considering the estimate $\hat{\boldsymbol{x}}$ as $\hat{d}_i=\|\hat{\boldsymbol{x}}-\boldsymbol{a}_i\|_2$ and form a new approximate weighting matrix. This approach can be
continued for a number of iterations, however, as we have observed through simulations, after two updatings, the estimation accuracy improves only slightly via additional iterations.

Another estimator based on a linear least squares (LLS) approach obtained in Appendix~\ref{sec:LLS} can be alternatively applied to the model in \eqref{eq:sq-range}.  Note that the algorithm derived in Appendix~\ref{sec:LLS}
is similar to the one proposed in \cite{RobustClock_Localization_2011}, except the correction technique introduced in this study.
As  will be observed in the simulations section the proposed approach in this section, i.e., SQ-LS,
has better performance than the LLS approach, especially for low number of reference nodes.

\subsection{A concave-convex procedure (CCCP)}
\begin{figure*}
 \centering
\includegraphics[width=160mm]{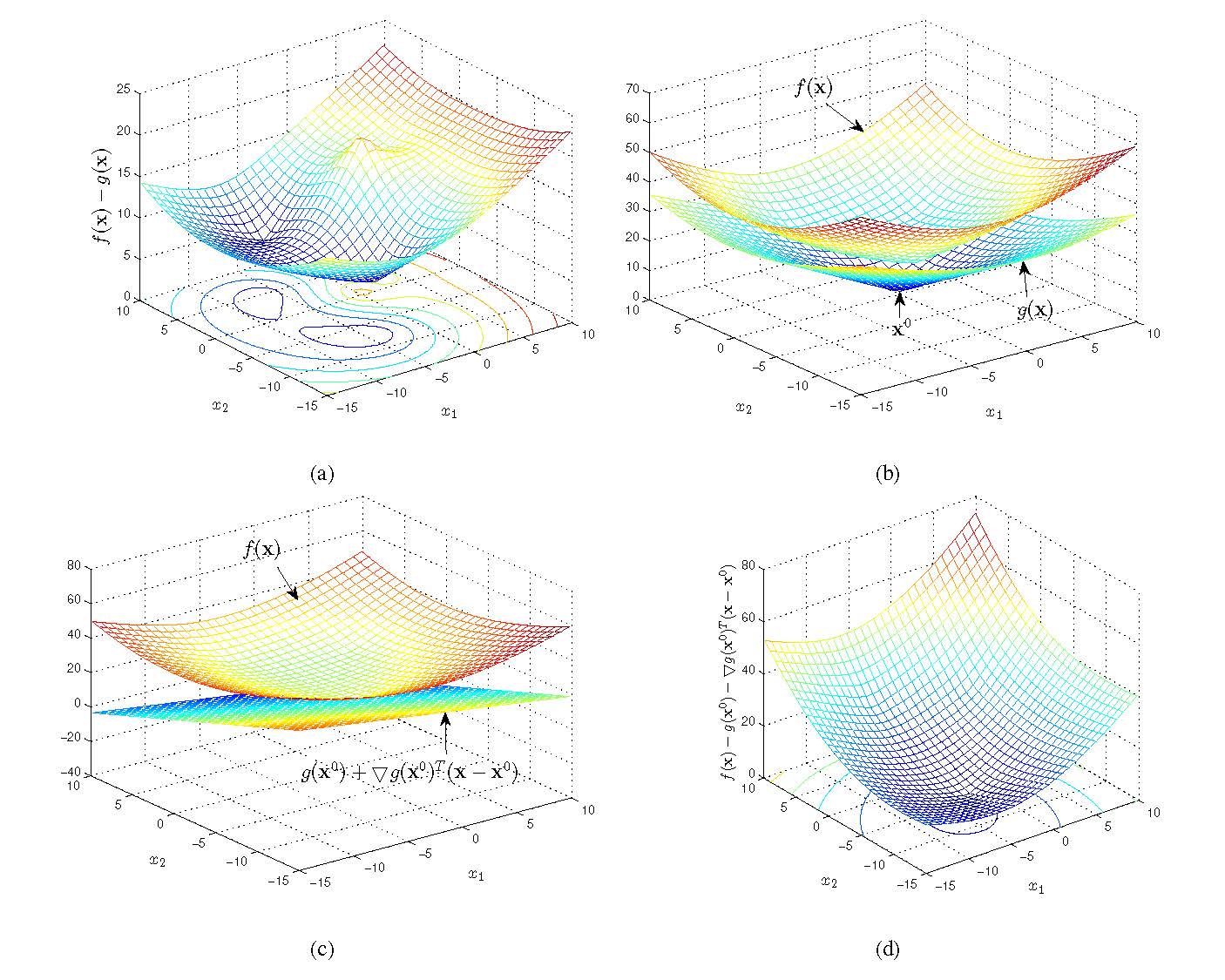}
\label{fig:4}
\caption{An example of the CCCP procedure. (a) A nonconvex function in the form of the difference of two convex functions and its contour,
 (b) separation of the nonconvex function into two convex functions, (c) the first order approximation of the second function, and (d)
 a convex approximation of the original nonconvex function at point $\boldsymbol{x}^0=[0~0]$.}
 \label{fig:cccp_example}
\end{figure*}

In this section, we take the $\ell_1$ norm minimization of residual errors into account and
propose a technique to solve the positioning problem.
Namely, based on \eqref{eq:sub_model}, we consider the following $\ell_1$ norm  minimization problem:
\begin{align}
\label{eq:l1minimization1}
\mathop{\text{minimize}}\limits_{\boldsymbol{x}\in\mathbb{R}^2;~\alpha\in\mathbb{R}_+}~\|\boldsymbol{r}\|_{1}
\end{align}
where $\boldsymbol{r}=[r^1_1 \ldots r^1_{N} \ldots r^K_1 \ldots r^K_{N}]^T$ with $r^k_i=z^k_{i}\alpha-\hat{T}_i^k/2-d_i/c$.
Note that for high signal-to-noise ratios (low standard deviations of noise), the $\ell_2$ and $\ell_1$ minimization approaches
have similar performance~\cite{boyd_convex}. Moreover, the $\ell_1$ norm minimization in \eqref{eq:l1minimization1} is  robust
against outliers~\cite{boyd_convex}.
The optimization problem
in~\eqref{eq:l1minimization1} can be written (in the epigraph form) as~\cite{boyd_convex,DC_gholami_2013,Mohammad_thesis}
\begin{align}
\label{eq:l1minimization2}
\mathop{\text{minimize}}\limits_{\boldsymbol{x}\in\mathbb{R}^2; \alpha\in\mathbb{R}_+; \boldsymbol{t}\in\mathbb{R}^N}~&\mathop{\sum}\limits_{k=1}^K\mathop{\sum}\limits_{i=1}^{N} t^k_i\nonumber\\
\text{subject to}~~&z^k_{i}\alpha-\hat{T}_i^k/2-d_i/c\leq t^k_i\nonumber\\
&z^k_{i}\alpha-\hat{T}_i^k/2-d_i/c\geq -t^k_i.
\end{align}
The nonconvex problem in~\eqref{eq:l1minimization2} is reminiscent of a well-known nonconvex problem
called difference of convex functions programming (DCP) \cite{CCCP_Horst_99}. The general form of DCP is as follows:
\begin{align}
\label{eq:CCCP}
&\mathop{\text{minimize}}\limits_{\boldsymbol{x}}~f_0(\boldsymbol{x})-g_0(\boldsymbol{x})\nonumber\\
&\text{subject to}~w_i(\boldsymbol{x})-g_i(\boldsymbol{x})\leq 0,~i=1,\ldots,M
\end{align}
where $w_i(\boldsymbol{x})$ and $g_i(\boldsymbol{x})$ are both smooth convex functions for $i=1,\ldots,M$.
A method to solve~\eqref{eq:CCCP} is to sequentially solve the problem. That is, we first
approximate the concave function $(-g_i(\boldsymbol{x}))$ with a convex one by an affine approximation.
Let us consider a point $\boldsymbol{x}^j$
in the domain of the problem in \eqref{eq:CCCP}, linearize the concave function around $\boldsymbol{x}^j$
and write the optimization problem in \eqref{eq:CCCP} as
\begin{align}
\label{eq:CCCP_convex}
&\mathop{\text{minimize}}\limits_{\boldsymbol{x}}~f_0(\boldsymbol{x})-g_0(\boldsymbol{x}^j)-\bigtriangledown g_0(\boldsymbol{x}^j)^T(\boldsymbol{x}-\boldsymbol{x}^j)\nonumber\\
&\text{subject to}~~w_i(\boldsymbol{x})-g_i(\boldsymbol{x}^j)-\bigtriangledown g_i(\boldsymbol{x}^j)^T(\boldsymbol{x}-\boldsymbol{x}^j)\leq 0.
\end{align}
The convex problem in~\eqref{eq:CCCP_convex} can now be solved efficiently. Denote the solution of \eqref{eq:CCCP_convex} as $\boldsymbol{x}^{j+1}$.  Next
we go for further improving the solution by convexifing \eqref{eq:CCCP} for the new point  $\boldsymbol{x}^{j+1}$ similar to the procedure employed
for $\boldsymbol{x}^{j}$. This sequential programming procedure, called concave-convex programming (CCCP),  continues for a number of iterations.
The convergence of the CCCP to a stationary point has been shown in the literature, e.g.,~\cite{CCCP_Horst_99,CCCP_2011} and references therein.
Fig.\,\ref{fig:cccp_example} shows an example of the CCCP approach for a DCP, where $f(\boldsymbol{x})$ and $g(\boldsymbol{x})$ are as follows:
\begin{align}
&f(\boldsymbol{x})=\|\boldsymbol{x}-\boldsymbol{a}_1\|_2+\|\boldsymbol{x}-\boldsymbol{a}_2\|_2+\|\boldsymbol{x}-\boldsymbol{a}_3\|_2,\nonumber\\
&g(\boldsymbol{x})=\|2\boldsymbol{x}-\boldsymbol{a}_4\|_2,
\end{align}
with $\boldsymbol{a}_1=[3~1.7],~\boldsymbol{a}_1=[-6~5],~\boldsymbol{a}_3=[-4~-7]$, and $\boldsymbol{a}_4=[0.5~1]$. In this figure, the original nonconvex problem is transferred to
a convex problem using a linear approximation of the nonconvex problem around $\boldsymbol{x}^0=[0~0]$.

Applying the CCCP technique to the problem in \eqref{eq:l1minimization2}, we get the following optimization problem:
\begin{align}
\label{eq:l1minimization_cccp}
\mathop{\text{minimize}}\limits_{\boldsymbol{x}\in\mathbb{R}^2;\alpha\in\mathbb{R}_+;\boldsymbol{t}\in\mathbb{R}^N}~&\mathop{\sum}\limits_{k=1}^{K}\mathop{\sum}\limits_{i=1}^{N} t^k_i\nonumber\\
\text{subject to}~~&z^k_i\alpha-\boldsymbol{h}_{i,j}^T\boldsymbol{x}-b^j_{i,k}-t^k_i\leq 0\nonumber\\
& \frac{1}{c}\|\boldsymbol{x}-\boldsymbol{a}_i\|_2-z^k_i\alpha+\frac{\hat{T}_i^k}{2}-t^k_i\leq 0
\end{align}
where $\boldsymbol{h}_{i,j}=(\boldsymbol{x}^j-\boldsymbol{a}_{i})/(cd(\boldsymbol{a}_{i},\boldsymbol{x}^j))$ and
$b^j_{i,k}={\hat{T}_i^k}/{2}-\boldsymbol{h}_{i,j}^T\boldsymbol{x}^j+d(\boldsymbol{a}_{i},\boldsymbol{x}^j)/c$.
The optimization problem in \eqref{eq:l1minimization_cccp},
which is called second order cone programming (SOCP), can be efficiently solved. We call the corresponding CCCP
as CCCP-SOCP.

Note that if $g_i(\boldsymbol{x})$ is not differentiable at $\boldsymbol{x}^j$, we can replace the $\bigtriangledown g_i(\boldsymbol{x}^j)$ term
by a subgradient\footnote{Let $\mathcal{D}$ be a nonempty set in $\mathbb{R}^n$. A vector $\boldsymbol{g}\in\mathbb{R}^n$ is a subgradient of a function $f:\mathcal{D}\rightarrow \mathbb{R}$ at $\boldsymbol{x}\in \mathcal{D}$ if
${f(\boldsymbol{y})\geq f(\boldsymbol{x})+\boldsymbol{g}^T(\boldsymbol{y}-\boldsymbol{x})}$ for all $\boldsymbol{y}\in\mathcal{D}$~\cite{Censor_97}.} of $g_i(\boldsymbol{x})$ at $\boldsymbol{x}^j$.

\section{ Complexity analysis}
\label{Sec:complexiy}
In this section, we study the complexity of the proposed techniques in terms of floating point operations (\emph{flops}) and also running time in Matlab.
We compare the complexity of the MLE, LLS, SQ-LS, and CCCP-SOCP.
To compute the complexity of the MLE, we assume that a good initial point is available, and an iterative algorithm such as the Gauss-Newton (GN) method converges to the global minimum after a number of iterations. Of course, finding a good initial point for the MLE is a challenging problem and this study also aims to tackle it. For the problem at hand the complexity of the MLE for every Newton step can be computed as $O(K^2N^3))$. For the AMLE, the complexity for every Newton step can be computed as $O((KN)^2)$.  The corresponding LLS needs
an order of $O(KN)$ to implement.
For the SQ-LS, we need to use a bisection search to solve \eqref{eq:gamm}, which is the most complex part of the algorithm.
Suppose the bisection search takes $k_{\mathrm{sq}}$ steps, then the total cost of the the proposed approach can be approximated as
$O(k_{\mathrm{sq}}KN)$.
In the simulations, we have observed that the bisection search algorithm usually takes 20 to 30 iterations to find the optimal value of $\gamma$.
Note that we need to run the LLS and SQ-LS twice. Thus the corresponding complexities are increased by a factor of two.
Finally, the complexity of the CCCP-SOCP can be computed as follows.
Consider a general form of the SOCP problem as
\begin{align}
\label{eq:SOCP_general}
\mathop{\mathrm{minimize}}\limits_{\boldsymbol{x}\in\mathbb{R}^n} ~&\boldsymbol{c}^T\boldsymbol{x}\nonumber\\
\mathrm{subject~ to}&~\|\boldsymbol{A}_i\boldsymbol{x}+\boldsymbol{b}_i\|_2\leq \boldsymbol{c}_i^T\boldsymbol{x}+d_i,~i=1,\ldots,m,\nonumber\\
& \|\boldsymbol{x}\|_2\leq R
\end{align}
where  $\boldsymbol{A}_i\in\mathbb{R}^{k_i\times n}, \boldsymbol{b}_i\in\mathbb{R}^{k_i}$, and $d_i\in\mathbb{R}$.
Note that the constraint on the norm of $\boldsymbol{x}$ ensures the strong convexity of the centering problem in the barrier approach~\cite{boyd_convex}. The worst-case complexity of the problem in \eqref{eq:SOCP_general} can be computed as~$O((1+m)^{1/2}n(n^2+m+\sum_{i=1}^m k_i^2)\log{1/\epsilon})$ \cite{Nemirovski_lecturenote_2012}, where $\epsilon$ is an accuracy tolerance in solving the problem.

The complexity of the CCCP-SOCP for every estimate
$\boldsymbol{x}^j$ can now be approximated as~
\[O((KN)^{3.5}\log{1/\epsilon}).\] As mentioned before we need to solve the problem in $k_{\mathrm{cccp}}$ steps, hence the total cost is
$O(k_{\mathrm{cccp}} (KN)^{3.5}\log{1/\epsilon})$. As we observe, a small number of updatings, usually three, $k_{\mathrm{cccp}}=3$, is enough to obtain the solution.
Table\, \ref{tab:cost} summarizes the complexity of the different approaches.

\begin{table*}
\footnotesize
\centering
\caption{Complexity of different approaches.}
    \begin{tabular}{ | l | l | l |p{4cm} |}
    \hline
      Method &  Complexity \\   \hline\hline
     MLE (GN initialized with good initial points)&  $O(k_{\mathrm{mle}} K^2N^3)$ \\ \hline
     AMLE (GN initialized with good  initial points)&  $O(k_{\mathrm{amle}} (KN)^2)$ \\ \hline
   CCCP-SOCP & $O(k_{\mathrm{cccp}} (KN)^{3.5}\log{1/\epsilon})$ \\ \hline
   LLS&  $O(KN)$ \\ \hline
   SQ-LS&$O(k_{\mathrm{sq}}KN)$\\ \hline
    \end{tabular}
   \label{tab:cost}
\end{table*}

We have also measured the average running time of different algorithms for a network consisting of 6 reference nodes as considered in Section \ref{sec:simulation}. In the simulations, we set $K=2$ and $\sigma_i=\gamma_i=10$.
The algorithms have been implemented  in Matlab on a MacBook Pro (Processor  2.3 GHz Intel Core i7, Memory  8 GB 1600 MHz DDR3). The MLEs are implemented
by Matlab function \emph{fminsearch} initialized with the true values of the target position, the clock skew, and turn-around times. The CCCP-SOCP is implemented by the \emph{CVX} toolbox~\cite{cvx}. We use three updatings to get an estimate.

We run the algorithms for 500 realizations of the network and compute the average running time in ms. The results are shown in Table\,\ref{tab:cost2}.
Considering the complexity analysis and average running time in Tables\,\ref{tab:cost} and \ref{tab:cost2}, respectively, we can conclude that the proposed approach has reasonable complexity and running time.
Although CCCP-SOCP takes a longer amount of time than MLE, it does not need a good initial point. While for the MLE with an arbitrary initial point, the algorithm may converge to a local minimum resulting in a large position error.
As we will see in the next section, the CCCP-SOCP outperforms both the LLS and SQ-LS approaches in terms of the root-mean-squared-error.

\begin{table*}
\footnotesize
\centering
{\caption{ Running time of Different Algorithms.}
    \begin{tabular}{ | l | l | l |p{4cm} |}
    \hline
    Method &   Time (ms)\\ \hline \hline
     AMLE (GN initialized with true values) &  32
 \\ \hline
   MLE (GN initialized with true values) &  317
 \\ \hline
     LLS  &  0.74 \\ \hline
     SQ-LS&6.2\\ \hline
   CCCP-SOCP&976\\ \hline
    \end{tabular}
   \label{tab:cost2}}
\end{table*}

\section{Numerical results}
\label{sec:simulation}
In this section, we evaluate the performance of the proposed approaches through computer simulations.
We consider a $1600$ m by $1600$ m area and a number of reference nodes that are located at fixed positions
$\boldsymbol{a}_1=[800~800],~\boldsymbol{a}_2=[800~-800],~\boldsymbol{a}_3=[-800~800],~\boldsymbol{a}_4=[-800~-800],~\boldsymbol{a}_5=[800~800],~\boldsymbol{a}_6=[0~800],~\boldsymbol{a}_7=[-800~0]$, and $\boldsymbol{a}_8=[0~-800]$.
In the simulations, we pick the first $N$ reference nodes, i.e., $\boldsymbol{a}_1,\ldots,\boldsymbol{a}_N$. One target node is randomly distributed inside
the area. The turn-around time is set to $0.001$ ms. The clock skew is assumed to be unknown and is set to $100$ PPM, i.e., $w=1.0001$.
Such a value for clock skew is common for a practical oscillator. For example for a center carrier frequency $f_c=100$ MHZ and a frequency offset $\Delta f=10$ KHZ, we can write
\begin{align}
T=\frac{1}{f_c\mp \Delta f}\approx \frac{1}{f_c}(1\pm \frac{\Delta f}{f_c})=\frac{1}{f_c}(1\pm 0.0001).
\end{align}
We compare the proposed techniques (CCCP-SOCP and SQ-LS) with the MLE and AMLE in \eqref{eq:MLE1} and \eqref{eq:MLE2}, respectively, (which are implemented by Matlab function \emph{fminsearch}~\cite{Mathworks_2012} initialized with
the true values of the target location, turn-around times, and clock skew),
the LLS derived in Appendix~\ref{sec:LLS}, and the Cram\'er-Rao lower bound (CRLB) as derived in Appendix~\ref{sec:CRLB}. In the simulations, we assume that $\sigma_i=\gamma_i=\sigma,~i=1,\ldots,N$.
We randomly initialize the CCCP-SOCP inside the network and we also set
$K_{\mathrm{cccp}}=3$.
 To simulate the range measurements and estimates of turn around times, we use
models \eqref{eq:turn} and \eqref{eq:3}, respectively.
To implement the bisection search, we consider an interval defined by $I_{\text{lower}}$ and $I_{\text{upper}}$ and investigate if the zero crossing of
$\phi(\mu)$ in \eqref{eq:gamm}
occurs in the interval. To check if the solution lies in the interval, we simply check the sign of $\phi(\mu)$ at $I_{\text{lower}}$ and $I_{\text{upper}}$.  No change in sign means that the solution lies outside of the current interval.
For initialization, we set $I_{\text{lower}}=-1/\mu_1$ and $I_{\text{upper}}=1/\mu_1$. If the solution of $\phi(\mu)=0$
is not found in the interval, we change the interval as
$I'_{\text{lower}}=I_{\text{upper}}$ and $I'_{\text{upper}}=10I_{\text{upper}}$.
If the solution lies in an interval, we bisect the interval and investigate which subinterval contains the solution.

\begin{figure}
 \centering
 \includegraphics[width=160mm]{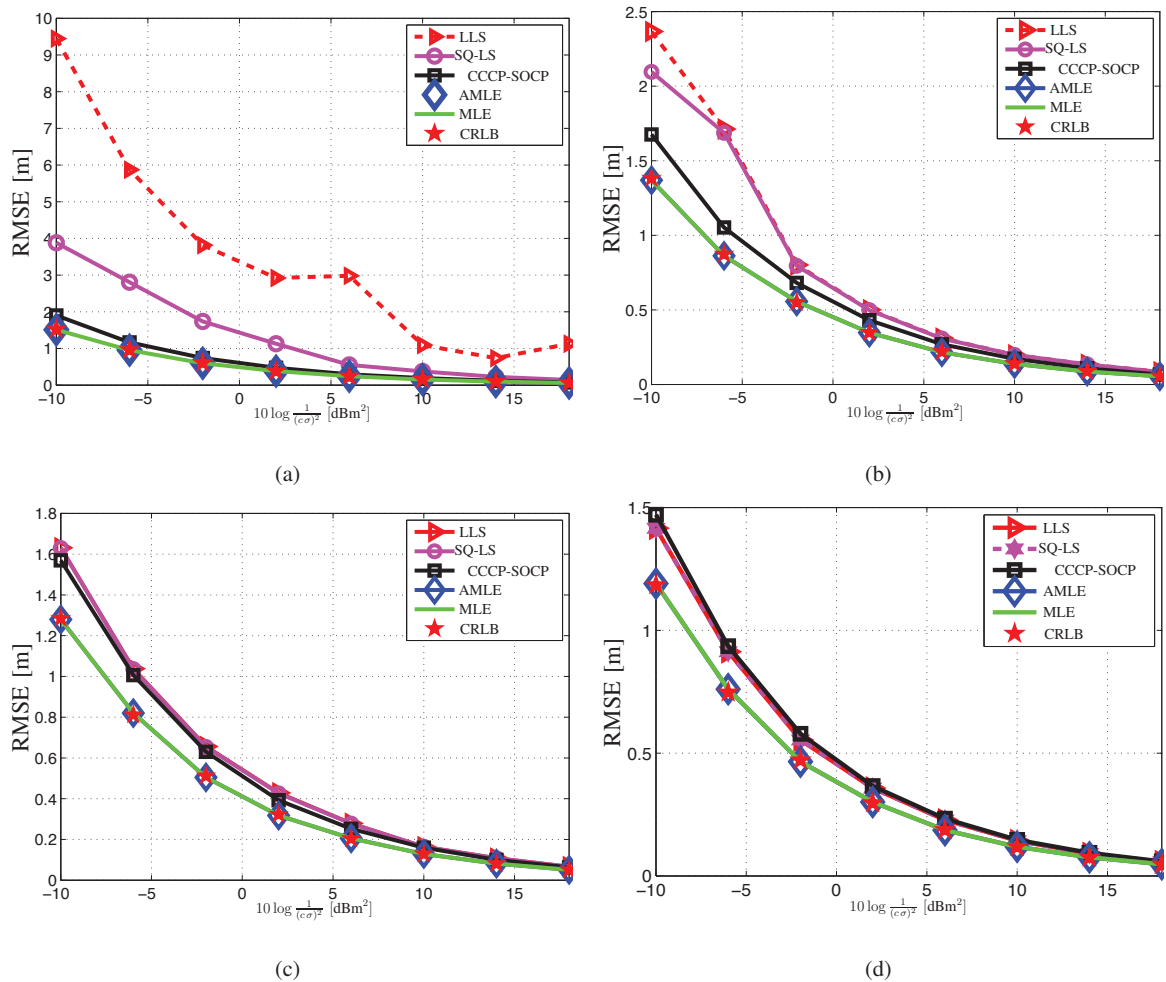}
\label{fig:8ref}
\caption{The RMSE of difference approaches for $K=2$ for (a) five reference nodes,
 (b) six reference nodes, (c) seven reference nodes, and (d) eight reference nodes.}
 \label{fig:rmse}
\end{figure}

Fig.\,\ref{fig:rmse} shows the root-mean-squared-errors (RMSEs) of location estimates for different approaches
for various numbers of reference nodes. In the simulations, we set $K=2$.
It is observed that the proposed approach, CCCP-SOCP, achieves good performance very close to the optimal estimator MLE and
the CRLB, especially for low number of reference nodes and high signal-to-noise ratios.
From the figure, it is noted that the SQ-LS performance is worse than CCCP-SOCP, but better than LLS, especially for a low number of reference nodes. As the number of reference nodes increases, the  LLS and SQ-LS show similar performance.
It is also observed that for large numbers of reference nodes, the least squares based approach shows better performance compared to the CCCP-SOCP approach for the low standard deviation of noise.

Next, we study the effects of NLOS measurements on the performance of estimators.
We assume that a range measurement can be affected by NLOS errors with probability 0.2. For every
NLOS measurement, we add a uniform noise to the measurements as follows:
\begin{align}
\label{eq:measurement}
z^k_i=w\left(\frac{d_i}{c}+\frac{T_i}{2}\right)+u^k_i+\frac{n^k_i}{2}
\end{align}
where we assume that $u^k_i\sim \mathcal{U}[0,5/c]$.
The uniform distribution is commonly used to model NLOS error, e.g., \cite{Mats_thesis_2008,Urruela_2006,Gholami_Eurasip_2011}.

\begin{figure}
 \centering
  \includegraphics[width=160mm]{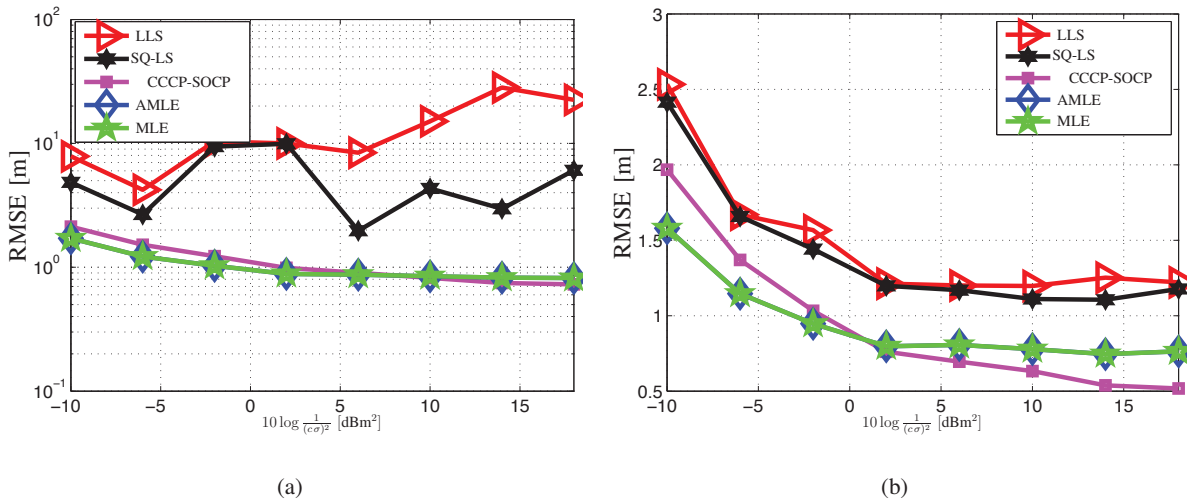}
\label{fig:7ref}
\caption{The RMSE of difference approaches for NLOS conditions ($K=2$) for (a) five reference nodes and
 (b) six reference nodes.}
 \label{fig:rmse_nlos}
\end{figure}

Fig.\,\ref{fig:rmse_nlos} depicts the performance of different approaches in NLOS conditions for $K=2$.
It is observed that the CCCP-SOCP achieves high performance compared to the other approaches, especially for low standard deviations of noise,  and it is robust against outliers as expected.  For small $\sigma$, the dominant perturbation is outlier disturbance and consequently the MLE derived in this study is not optimal, explaining why the MLE is worse than the CCCP-SOCP approach. For large standard deviations of noise, which indicates the Gaussian measurement noise is dominant, the CCCP-SOCP seems to outperform the MLE.
This can be explained by the fact that the MLE is only guaranteed to be asymptotically optimal, i.e., for low noise standard deviation or large number of measurements.
Note that we have employed the MLE computed in \eqref{eq:MLE1} and \eqref{eq:MLE2} to study their robustness against NLOS conditions.
It may be possible to derive an MLE to deal with NLOS measurements if the distribution of outliers is known.

\begin{figure}
 \centering
\includegraphics[width=160mm]{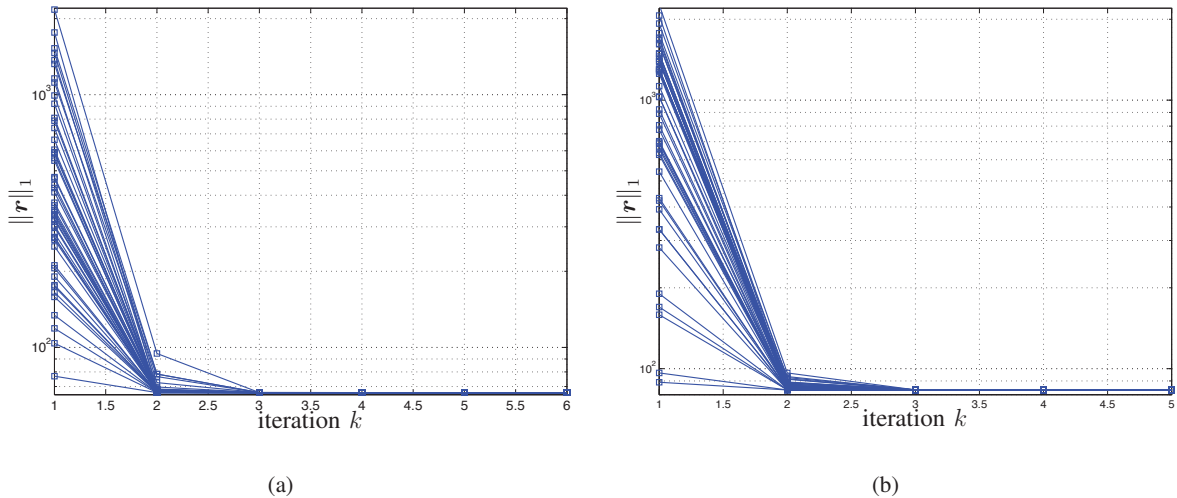}
\label{fig:6ref}
\caption{Convergence of proposed approaches for 50 random initializations for $c\sigma=10$ and $K=2$ for (a) 6 reference nodes,
 (b) 8 reference nodes.}
 \label{fig:convergence}
\end{figure}

Finally, we study the convergence of CCCP-SOCP through simulations.
Fig.\,\ref{fig:convergence} depicts the convergence speed of the proposed approach for 50 random initializations. In the simulations, we set $K=2$.
For every estimate given by CCCP-SOCP, we compute the residual $\|\boldsymbol{r}\|_1$, where $\boldsymbol{r}$ is given by \eqref{eq:l1minimization1}.
It is observed that the CCCP-SOCP approach converges very fast, approximately in three sequential updatings.

\section{Conclusions}
\label{sec:conclude}
In this paper, we have studied TW-TOA based positioning in an asynchronous network.
Since the optimal ML estimator is highly nonconvex and difficult to solve,
we have obtained two efficient  suboptimal estimators for the problem under some approximations and conditions.
The first method is based on squared-range least squares which can be solved exactly under
mild conditions. The second approach is derived by replacing the $\ell_2$ norm minimization
of residuals by an $\ell_1$ norm minimization, which in turn can be formulated as difference of convex programming (DCP).
We have used a concave-convex procedure to solve the resulting DCP. Simulation results show the high performance of the proposed
techniques, especially the DCP approach. It has also been observed through simulations that the DCP approach is robust against NLOS errors.


\begin{appendices}
\section{Linear least squares (LLS)}
\label{sec:LLS}
In this section we obtain an LLS estimator similar to
\cite{Ho_2009_Succesive_Asymptotically,Gholami_Coop_2012}.
We consider the following linear model (originated from \eqref{eq:sq-range}):
\begin{align}
\label{eq:sec}
\boldsymbol{b}=\boldsymbol{A}\boldsymbol{y}+\boldsymbol{\nu},
\end{align}
where $\boldsymbol{\nu}=[\nu^1_1~\ldots~\nu^1_N \ldots \nu^K_1~\ldots~\nu^K_N]^T$, $\boldsymbol{A}$,  $\boldsymbol{b}$, and $\boldsymbol{y}$ are given in \eqref{eq:linearMod_parameters}.
We assume that $\boldsymbol{A}$ has full column rank. A necessary condition for this is that
$KN\geq 5$.

The unconstrained weighted least squares solution to \eqref{eq:sec} is given by~\cite{Kay_93}
  \begin{align}
  \label{eq:firstestimate}
 \hat{\boldsymbol{y}}=(\boldsymbol{A}^T\boldsymbol{W}\boldsymbol{A})^{-1}\boldsymbol{A}^T\boldsymbol{W}\boldsymbol{b}.
 \end{align}
 where $\boldsymbol{W}$ is as in \eqref{eq:linearMod_parameters}.
 The covariance matrix of $\hat{\boldsymbol{y}}$ can be computed as
\begin{align}
\boldsymbol{C}_{\hat{\boldsymbol{y}}}=(\boldsymbol{A}^T\boldsymbol{W}\boldsymbol{A})^{-1}.
\end{align}


Note that for a large network, matrix $\boldsymbol{A}$ can be
 ill-conditioned~\cite{Gholami_Coop_2012}.  Then, we can use a regularization technique to resolve the drawback in the least squares solution~\cite{boyd_convex,Gholami_Coop_2012}.

We can further improve the location estimate by applying a correction technique similar to \cite{Ho_2009_Succesive_Asymptotically,Gholami_Coop_2012}. We consider the following relations:
\begin{align}
\label{eq:est_elements}
&[\boldsymbol{y}]_1=\|x\|^2_2+\xi_1, \nonumber\\
&[\boldsymbol{y}]_4=\alpha^2+\xi_4,\nonumber\\
&[\boldsymbol{y}]_2=x_1+\xi_2, \nonumber\\
&[\boldsymbol{y}]_3=x_2+\xi_3, \nonumber\\
&[\boldsymbol{y}]_5=\alpha+\xi_5,
\end{align}
where $\boldsymbol{\xi}=[\xi_1~\ldots~\xi_5]^T$ is the estimation error. Assuming small estimation  errors, we take the squares of both sides of last three equations in \eqref{eq:est_elements} and obtain the following expressions:
\begin{align}
\label{eq:est_elements_2}
&[\boldsymbol{y}]^2_2\simeq x^2_1+2 x_1\xi_2, \nonumber\\
&[\boldsymbol{y}]^2_3\simeq x^2_2+2x_2 \xi_3, \nonumber\\
&[\boldsymbol{y}]^2_5\simeq\alpha^2+2\alpha\xi_5.
\end{align}
Based on \eqref{eq:est_elements} and \eqref{eq:est_elements_2}, we obtain a linear model as
\begin{align}
\label{eq:linearmod_2}
\boldsymbol{h}=\boldsymbol{B}\boldsymbol{\theta}+\boldsymbol{P}\boldsymbol{\xi},
\end{align}
where
\begin{align}
&\boldsymbol{B}=
\left[\begin{array}{ccc}1&1&1\\
1& 0& 0\\
0& 1& 0\\
0& 0& 1
\end{array}\right],\quad
\boldsymbol{P}=
\left[\begin{array}{ccccc}1&0&0&0&1\\
0&2x_1& 0&0&0\\
0& 0&2x_2& 0&0\\
0& 0& 0&2\alpha&0\\
\end{array}\right]\nonumber\\
&\boldsymbol{h}=
\left[\begin{array}{ccc}[\boldsymbol{y}]_1+[\boldsymbol{y}]_4\\
{[\boldsymbol{y}]^2_2}\\
{[\boldsymbol{y}]^2_3}\\
{[\boldsymbol{y}]^2_4}
\end{array}\right],\quad \boldsymbol{\theta}=[x^2_1~ x^2_2~ \alpha^2]^T.
\end{align}

The least squares solution to \eqref{eq:linearmod_2} is given by
\begin{align}
\hat{\boldsymbol{\theta}}=(\boldsymbol{B}^T\boldsymbol{C}^{-1}_{\hat{\boldsymbol{\theta}}}\boldsymbol{B})^{-1}
\boldsymbol{B}^T\boldsymbol{C}^{-1}_{\hat{\boldsymbol{\theta}}}\boldsymbol{h},
\end{align}
where the covariance matrix $\boldsymbol{C}_{\boldsymbol{\theta}}$ can be computed as
\begin{align}
\boldsymbol{C}_{\hat{\boldsymbol{\theta}}}=\boldsymbol{P}\boldsymbol{C}_{\hat{\boldsymbol{y}}}\boldsymbol{P}^T.
\end{align}
To compute the matrix $\boldsymbol{P}$, we use the estimate of $\hat{\boldsymbol{x}}$ obtained in \eqref{eq:firstestimate} instead of unknown vector~$\boldsymbol{x}$.

Finally the location estimate can be obtained as
\begin{align}
\label{eq:final_estimate}
\tilde{x}_i=\mathrm{sgn}([\boldsymbol{y}]_{i+1})\sqrt{|[\hat{\boldsymbol{\theta}}]_i|},\quad i=1,2,
\end{align}
where $\mathrm{sgn}$ denotes the signum function defined as
\begin{align}
\mathrm{sgn}(x) = \left\{ \begin{array}{ll}
         1 &\text{if}~ x \geq 0;\\
        -1 & \text{if}~ x < 0.\end{array} \right.
 \end{align}

The covariance matrix of the estimate in \eqref{eq:final_estimate} can be obtained similar to \cite{Gholami_Coop_2012}.

\section{Cram\'er-Rao Lower Bound (CRLB)}
\label{sec:CRLB}
Considering the measurement vector in \eqref{eq:meas} with mean
$\boldsymbol{\mu}_K=\boldsymbol{1}_K\otimes\boldsymbol{\mu}$
and covariance matrix $\boldsymbol{C}_K=\boldsymbol{I}_K\otimes \boldsymbol{C}$
where
\begin{align}
\label{eq:mean_cov}
&\boldsymbol{\mu}=\left[f\left(\frac{d_1}{c}+\frac{T_1}{2}\right) \ldots f\left(\frac{d_N}{c}+\frac{T_N}{2}\right)~ T_1\ldots T_N\right]^T,\nonumber\\
&\boldsymbol{C}=\text{diag}\left(\frac{\sigma_1^2}{4},\ldots,\frac{\sigma_N^2}{4},\gamma_1^2,\ldots,\gamma_N^2\right),
\end{align}
the elements of the Fisher information matrix can be computed as~\cite[Ch. 3]{Kay_93}
\begin{align}
J_{nm}=[\boldsymbol{J}]_{nm}=\left[\frac{\partial{\boldsymbol{\mu}_K}}{\partial{\psi_n}}\right]^T\boldsymbol{C}_K^{-1}\left[\frac{\partial{\boldsymbol{\mu}_K}}{\partial{\psi_m}}\right],\quad n,m=1,2,\ldots, N+3,
\end{align}
where
\begin{align}
 \psi_n=
 \begin{cases}
 x_{n},&\mathrm{if}~n=1,2\\
 w,&\mathrm{if}~n=3\\
 T_n,&\mathrm{if}~n>3.
 \end{cases}~
\end{align}
From~\eqref{eq:mean_cov},
$\partial{\boldsymbol{\mu}_K}/\partial{\psi_n}$ can be obtained as
follows:
\begin{align}
&\left[\frac{\partial{\boldsymbol{\mu}_K}}{\partial{\psi_n}}\right]= \boldsymbol{1}_{K}\otimes\left[\frac{\partial{{\mu}_1}}{\partial{\psi_n}}
~\ldots~\frac{\partial{{\mu}_{N}}}{\partial{\psi_n}}\right]^T,\quad {n=1,2,\ldots,N+3},
\end{align}
where
\begin{align}
&\frac{\partial{\mu_{i}}}{\partial{\psi_n}}=
\begin{cases}
w\frac{x_{n}-a_{1,n}}{c\,d(\boldsymbol{a}_{i},\boldsymbol{x})}, &\text{if}~n=1,2 ,~ i\leq N\\
\frac{d_{i}}{c}+\frac{T_i}{2}, &\text{if}~n=3,~ i\leq N\\
0,&\text{if}~n=1, 2, \text{or}~ 3,~ i> N\\
\frac{w}{2},&\text{if}~n>3,~i\leq N\\
1,&\text{if}~n>3,~i> N.\\
\end{cases}
\end{align}
After some calculations, the entries of the Fisher information
matrix can be computed as follows:
\begin{align}
&J_{11}=4Kw^2\sum_{i=1}^N\left( \frac{x_{1}-a_{i,1}}{\sigma_i cd(\boldsymbol{a}_{i},\boldsymbol{x})}\right)^2,\nonumber\\
&J_{22}=4Kw^2\sum_{i=1}^N\left( \frac{x_{2}-a_{i,2}}{\sigma_i cd(\boldsymbol{a}_{i},\boldsymbol{x})}\right)^2,\nonumber\\
&J_{33}=4K\sum_{i=1}^N\left(\frac{d_i/c+T_i/2}{\sigma_i}\right)^2,\nonumber\\
&J_{jj}=K\left(\frac{2w^2}{\sigma_j^2}+\frac{1}{\gamma_j^2}\right), ~j>3\nonumber\\
&J_{12}=J_{21}=4Kw^2\sum_{i=1}^N\left( \frac{x_{1}-a_{i,1}}{\sigma_i c\,d(\boldsymbol{a}_{i},\boldsymbol{x})}\right)\left( \frac{x_{2}-a_{i,2}}{\sigma_i c\,d(\boldsymbol{a}_{i},\boldsymbol{x})}\right),\nonumber\\
&J_{13}=4Kw\sum_{i=1}^N\left( \frac{x_{1}-a_{i,1}}{\sigma_i cd(\boldsymbol{a}_{i},\boldsymbol{x})}\right)\left(\frac{d_i/c+T_i/2}{\sigma_i}\right),\nonumber\\
&J_{23}=4Kw\sum_{i=1}^N\left( \frac{x_2-a_{i,2}}{\sigma_i cd(\boldsymbol{a}_{i},\boldsymbol{x})}\right)\left(\frac{d_i/c+T_i/2}{\sigma_i}\right),\nonumber\\
&J_{j1}=J_{1j}=K\left(w\frac{x_{1}-a_{i,1}}{\sigma^2_i cd(\boldsymbol{a}_{i},\boldsymbol{x})}\right),\nonumber\\
&J_{j2}=J_{2j}=K\left(w\frac{x_{2}-a_{i,2}}{\sigma^2_i c\d(\boldsymbol{a}_{i},\boldsymbol{x})}\right),\nonumber\\
&J_{j3}=J_{3j}=4Kw\sum_{i=1}^N\left( \frac{x_{1}-a_{i,1}}{\sigma_i c\,d(\boldsymbol{a}_{i},
\boldsymbol{x})}\right)\left(\frac{d_i/c+T_i/2}{\sigma_i}\right),\nonumber\\
&J_{ij}=J_{ji}=0,\quad i\neq j,~ i,~j>3
\end{align}
The CRLB, which is a lower bound on the variance of any unbiased
estimator, is given as
\begin{align}
\label{eq:CRLBJointGen}
\mathrm{Var}(\hat{\psi_i})\ge
[\boldsymbol{J}^{-1}]_{i,i}\,.
\end{align}

%

\end{appendices}

\bibliographystyle{IEEEtran}

\bibliography{manuscript.bbl}

\newpage

\end{document}